\documentclass{aa}  
\usepackage{graphicx}
\usepackage[varg]{txfonts}
\usepackage[pdftex, colorlinks=true,linkcolor=blue,citecolor=blue,breaklinks=true]{hyperref}
\usepackage{natbib}

\usepackage{xspace}
\usepackage{multirow}

\begin{document} 

   \title{Mapping the neutral atomic hydrogen gas outflow in the restarted radio galaxy \object{\mbox{3C\,236}}}

   \author{R. Schulz\inst{1}
          \and
          R. Morganti\inst{1,2}
          \and
          K. Nyland\inst{3}
          \and
		  Z. Paragi\inst{4}
		  \and
		  E. K. Mahony\inst{5,6}
          \and
          T. Oosterloo\inst{1,2}
          }

   \institute{ASTRON, Netherlands Institute for Radio Astronomy, Postbus 2, 7990 AA, Dwingeloo, Netherlands\\
			  \email{schulz@astron.nl}
			 \and
			 Kapteyn Astronomical Institute, University of Groningen, PO Box 800, 9700 AV Groningen, The Netherlands 
             \and
             National Radio Astronomy Observatory, Charlottesville, VA 22903, USA
             \and
             Joint Institute for VLBI ERIC, Postbus 2, 7990 AA, Dwingeloo, Netherlands
             \and
             Sydney Institute for Astronomy, School of Physics A28, The University of Sydney, NSW 2006, Australia
			 \and
			ARC Centre of Excellence for All-Sky Astrophysics (CAASTRO), Australia
             }

   \date{--; --}

  \abstract
{The energetic feedback that is generated by radio jets in active galactic nuclei (AGNs) has been suggested to be able to produce fast outflows of atomic hydrogen (\ion{H}{I}) gas that can be studied in absorption at high spatial resolution. We have used the Very Large Array (VLA) and a global very-long-baseline-interferometry (VLBI) array to locate and study in detail the \ion{H}{I} outflow discovered with the Westerbork Synthesis Radio Telescope (WSRT) in the re-started radio galaxy 3C\,236. We confirm, from the VLA data, the presence of a blue-shifted wing of the \ion{H}{I} with a width of $\sim1000\mathrm{km\,s^{-1}}$. This \ion{H}{I} outflow is partially recovered by the VLBI observation. In particular, we detect four clouds with masses of $0.28\text{--}1.5\times 10^4M_\sun$ with VLBI that do not follow the regular rotation of most of the \ion{H}{I}. Three of these clouds are located, in projection, against the nuclear region on scales of $\lesssim 40\mathrm{\,pc}$, while the fourth is co-spatial to the south-east lobe at a projected distance of $\sim270\mathrm{\,pc}$. Their velocities are between $150$ and $640\mathrm{\,km\,s^{-1}}$ blue-shifted with respect to the velocity of the disk-related \ion{H}{I}. These findings suggest that the outflow is at least partly formed by clouds, as predicted by some numerical simulations and originates already in the inner (few tens of pc) region of the radio galaxy. Our results indicate that all of the outflow could consist of many clouds with perhaps comparable properties as the ones detected, distributed also at larger radii from the nucleus where the lower brightness of the lobe does not allow us to detect them. However, we cannot rule out the presence of a diffuse component of the outflow. The fact that 3C\,236 is a low excitation radio galaxy, makes it less likely that the optical AGN is able to produce strong radiative winds leaving the radio jet as the main driver for the \ion{H}{I} outflow.
  }
  

   \maketitle
%
\section{Introduction}
\label{sec:Intro}
	
	\begin{figure}[htpb]
		\centering
		\includegraphics[width=0.97\linewidth]{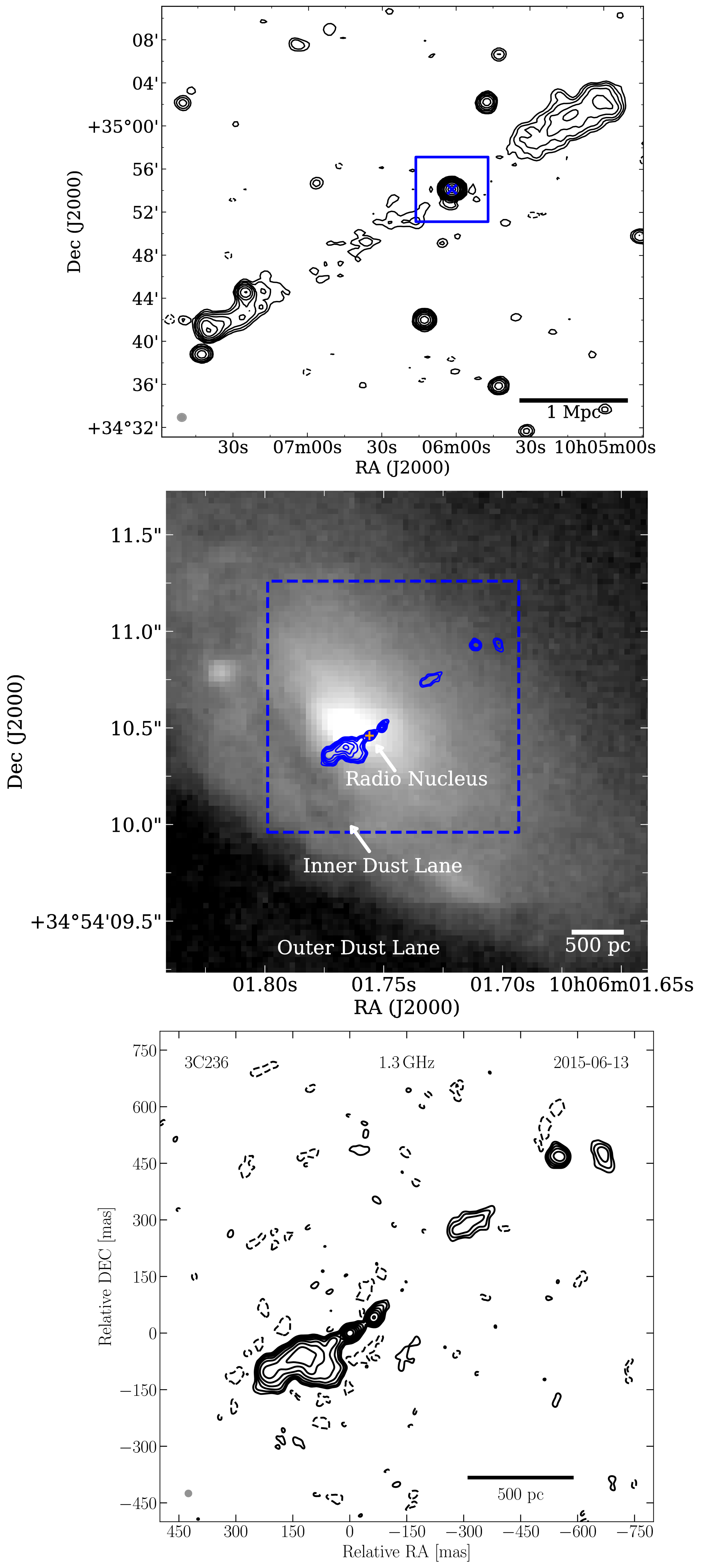}
		\caption{Left panel: 1.4\,GHz VLA NVSS image of the large-scale radio emission of \object{\mbox{3C\,236}} \citep{Condon1998}. The blue square and cross highlight the area covered by our VLA observation and the pointing of the VLBI observation, respectively. Middle panel: the gray-colored background shows a zoom-in into archival \textit{Hubble Space Telescope} V-band image (ACS/HRC/F555W, \citealt{ODea2001}). The blue contour lines trace the VLBI radio continuum emission starting for visibility at $5\times\sigma_\mathrm{noise,VLBI}$. The orange cross marks the position of the VLBI core from \cite{Taylor2001} to which our VLBI image was aligned to in this montage. The dashed lines mark the plot range of the VLBI image shown in the bottom panel. Bottom panel: VLBI image of obtained by our observation. The dashed and solid black contour lines trace negative and positive brightness starting from $3\times\sigma_\mathrm{noise,VLBI}$ and increasing logarithmically by a factor of 2.}
		\label{fig:Collage}%
	\end{figure}

	The evolution of galaxies is considered to be strongly linked to that of their central supermassive black holes (SMBH). The required feedback is commonly explained by a phase of enhanced activity related to the SMBH (e.g., \citealt{Heckman2014,Kormendy2013}). An active galactic nucleus (AGN) can affect the interstellar medium (ISM) by heating-up and expelling gas which hinders star formation and the accretion of matter onto the SMBH (e.g., \citealt{Silk1998,DiMatteo2005,Croton2006,McNamara2007}). Prominent observational signatures include outflows of ionized, molecular and atomic gas that have been associated with a number of AGN at a range of redshifts. The highest outflow rates have been determined for the cold ISM gas (molecular and atomic). Among the different possible drivers of these outflows are the radio jets launched in some AGN. The complex interplay between the AGN and the ISM requires detailed observational and theoretical studies of each phase of the outflowing gas (see reviews by \citealt{Veilleux2005,Fabian2012,Alexander2012,Wagner2016,Tadhunter2016,Harrison2017,Morganti2017b} and references therein).
	Here, we focus on the outflows of neutral atomic hydrogen (\ion{H}{I}) gas which have been observed in absorption in a number of radio sources with different radio power (e.g., \citealt{Morganti1998,Morganti2005,Morganti2013,Morganti2016,Oosterloo2000,Mahony2013,Gereb2015,Allison2015}). Some of these objects host young or re-started AGN where the central radio source shows characteristics of a compact steep spectrum (CSS) object. This provides valuable insight into the evolution of radio galaxies, because CSS sources are considered to be the younger counterparts of the much larger Fanaroff-Riley type radio galaxies (e.g., \citealt{ODea1998,Kunert-Bajraszewska2010,Orienti2016}). The radio continuum is commonly a few kpc or less in size which limits the spatial scales on which the \ion{H}{I} outflow can be observed. In the two radio galaxies 3C\,305 and 3C\,295, the outflows were found on kpc scales \citep{Morganti2005b,Mahony2013}. However, in most cases sub-arcsecond angular resolution is needed in order to locate the outflow and trace its structure.
	
	The angular resolution can be achieved by very long baseline interferometry (VLBI) which has been used to study the associated \ion{H}{I} gas in absorption in various radio sources (e.g., \citealt{Carilli1998,Peck1999,Peck2001,Vermeulen2003,Vermeulen2006,Struve2010,Struve2012,Araya2010}). The first detection of a broad \ion{H}{I}  outflow with VLBI was reported by \cite{Oosterloo2000} in the Seyfert 2 galaxy \object{\mbox{IC\,5063}}. Of particular relevance for our study was the succesful imaging and mapping of the \ion{H}{I} outflow in the young restarted radio galaxy \object{\mbox{4C\,12.50}} by \cite{Morganti2013}.
	The broad bandwidth and high sensitivity of the VLBI observation revealed the outflow in this source as an extended cloud co-spatial with the southern extent of the radio continuum emission. This is offset to the \ion{H}{I} gas at the systemic velocity which is located north of the nucleus. The study determined a mass of the cloud of up to 10$^5$ solar masses (M$_\sun$) and a mass outflow rate of $16\mathrm{-}29\mathrm{\,M_\sun\,yr^{-1}}$. A comparison with an unresolved \ion{H}{I} absorption spectrum obtained with the Westerbork Synthesis Radio Telescope (WSRT) showed that all of the absorption was recovered by VLBI. The results provide the strongest evidence for this type of radio AGN so far that the \ion{H}{I} outflow is driven by the jet. 
	
	Based on these results, we performed VLBI observations of \ion{H}{I} in absorption of a small sample comprising \object{\mbox{3C\,236}}, \object{\mbox{3C\,293}}, and \object{\mbox{4C\,52.37}}. This initial work will pave the way for future VLBI observations of \ion{H}{I} gas in a larger sample selected from the WSRT \ion{H}{I} absorption survey \citep{Gereb2015,Maccagni2017}.
	
	In this paper, we focus on \object{\mbox{3C\,236}} at a redshift of $z=0.1005$ \citep{Hill1996} which is one of the largest known radio galaxies extending about 4.5\,Mpc \citep{Willis1974,Barthel1985,Schilizzi2001}. 
	This source represents a re-started AGN, i.e., it exhibits signs of different stages of AGN activity. The large scale morphology (top panel in Fig. \ref{fig:Collage}) stems from a previous cycle of activity compared to the CSS-type radio source in its inner 2\,kpc region which is the result of the most recent cycle. The inner radio emission has a dynamical time scale consistent with the age of the young star formation region \citep{ODea2001,Schilizzi2001,Tremblay2010}.

	The host galaxy of \object{\mbox{3C\,236}} features a large outer and a smaller inner dust lane which are slightly offset in position angle (PA) with respect to each other \citep{ODea2001,Schilizzi2001,Labiano2013}. The inner dust lane has a PA of $\sim30\degr$ which is almost perpendicular to the sub-kpc scale radio jet. VLBI observations by \cite{Schilizzi2001} showed that the jet is oriented in north-west direction extending from the brightest feature which is synchrotron self-absorbed and thus likely to be the core region. The south-east lobe produced by the counter jet is positionally coincident with parts of the inner dust lane and its morphology is considered to be partially a result of jet-ISM interaction. The background image in the middle panel of Fig. \ref{fig:Collage} shows a zoom-in of the inner dust lane overlayed by the brightness distribution of the radio source as obtained in this paper (bottom panel of Fig. \ref{fig:Collage}, see Sect. \ref{sec:Results}).
	
	Low-resolution \ion{H}{I} absorption spectra reveal a deep narrow absorption feature near the systemic velocity \citep{vanGorkom1989} and a broad (up to 1000\,km\,s$^{-1}$) shallow blue wing corresponding to a mass outflow rate of $\sim 47\mathrm{M_\sun\,yr^{-1}}$ \citep{Morganti2005}. 
	Based on \ion{H}{I} VLBI observations by \cite{Struve2012}, the narrow component has been intepreted as \ion{H}{I} gas located in a regular rotating disk which is co-spatial to the south-east lobe about 250\,mas from the nucleus. Because of bandwidth limitations the VLBI data was not able to cover the velocity range of the outflow. 
	
	The optical AGN has been classified as a low-excitation radio galaxy (LERG, \citealt{Buttiglione2010,Best2012}) which makes it less likely that strong quasar or starbust-driven winds are the origin of the outflow, but rather the jets. There are also signs of an outflow of ionized gas \citep{Labiano2013}. The cold (CO) and warm (H$_2$) molecular gas have only been detected in a disk-like geometry aligned with position angle of the inner dust lane, but the latter has a significant turbulent component \citep{Nesvadba2011,Labiano2013}. 

	This paper presents new VLBI observations with a larger bandwidth than previous data to localise the \ion{H}{I} outflow with respect to the radio jet and constrain its properties in combination with new lower resolution Very Large Array (VLA) data. It is structured as follows: in Sect. \ref{sec:Obs} we present the data and subsequent calibration procedure. This is followed by a presentation of our results in Sect. \ref{sec:Results} and discussion in Sect. \ref{sec:Discussion}. We end the paper with our conclusions and a summary in Sect. \ref{sec:Summary}. 
	
	Throughout this paper, we adopt a standard $\Lambda$CDM-cosmology ($H_0 = 70 \mathrm{\,km\,s^{-1}\,Mpc^{-1}}$, $\Omega_m=0.3$, $\Omega_\lambda=0.7$) based on which 1.0\,mas corresponds to about 1.8\,pc for \object{\mbox{3C\,236}}. It is important to point out that a range of values are available for the systemic velocity $v_\mathrm{sys}$ (see also \cite{Struve2012}). \cite{Labiano2013} determined $v_\mathrm{sys}^\mathrm{CO} \approx 29761\mathrm{\,km\,s^{-1}}$ based on the CO spectrum, though the spectral setup prohibited sampling of the continuum emission at low velocities limiting the Gaussian fit to the spectrum. Nevertheless, this value is close to the SDSS value of $v_\mathrm{sys}^\mathrm{SDSS} \approx 29740\mathrm{\,km\,s^{-1}}$. \cite{Struve2012} reported a value of $v_\mathrm{sys}^\mathrm{\ion{H}{I}} \approx 29820\mathrm{\,km\,s^{-1}}$ assuming that parts of the observed \ion{H}{I} is constrained within a disk. However, the \ion{H}{I} was detected in absorption which makes it difficult to measure the full extent of the disk (see also Sect. \ref{sec:Discussion:disk}). Therefore, we use $v_\mathrm{sys}^\mathrm{CO}$ throughout this paper as a reference value.

\begin{table*}[t!]
	\centering
	\caption[]{Properties of Observations}
	\label{tab:Data:Observation}
	\begin{tabular}{ccccccccccc}
		\hline
		Array\tablefootmark{a} & Code\tablefootmark{b} & Date & $\nu_\mathrm{obs}$\tablefootmark{c} & $T_\mathrm{obs}$\tablefootmark{d} & Pol.\tablefootmark{e} & Correlator Pass & IFs & BW\tablefootmark{f} & $N_\mathrm{ch}$\tablefootmark{g} & $\Delta \nu$\tablefootmark{h}\\
		& & & [GHz] & [min] &  & & [MHz] & & [kHz] \\
		\hline
		\hline
		\multirow{2}{*}{EVN+VLBA+Ar} & \multirow{2}{*}{GN002B} & \multirow{2}{*}{2015-06-13} & \multirow{2}{*}{1.293}& \multirow{2}{*}{500}  & \multirow{2}{*}{Dual} & continuum & 4 & 16 & 32 & 500\\
			&	&	&	&	&	& spectral-line & 1 & 16 & 512 & 31.25\\
		VLA & 11A-166 & 2011-08-10 & 1.283 & 40 & Dual & spectral-line & 1 & 16 & 256 & 62.5\\
		\hline
	\end{tabular}
	\tablefoot{ 
		\tablefoottext{a}{Array used for observation. EVN: Effelsberg (Germany), phased-up Westerbork (5 stations, Netherlands), Jodrell-Bank (United Kingdom), Onsala (Sweden); VLBA (USA): Los Alamos (NM), Kitt Peak (AZ), St. Croix (VI), Mauna Kea (HI), Hancock (NH), Brewster (WA), Fort Davis (TX), North Liberty (IA), Pie Town (NM), Owens Valley (CA); Ar: Arecibo (Puerto Rico).}
		\tablefoottext{b}{Experiment Code}
		\tablefoottext{c}{Observing frequency}
		\tablefoottext{d}{Observing time. For the VLBI experiment, this represents the total on source time of the whole array.}
		\tablefoottext{e}{Polarization: Dual refers to two polarization were used (LL and RR)}
		\tablefoottext{f}{Bandwidth (of each IF)}
		\tablefoottext{g}{Number of channels in a single band or IF}
		\tablefoottext{h}{Channel width in frequency}
	}
\end{table*}

\begin{table*}[t!]
	\centering
	\caption[]{Properties of Images}
	\label{tab:Data:Image}
	\begin{tabular}{ccccccc}
		\hline
		Data & $\sigma_\mathrm{noise}$\tablefootmark{a} & Beam\tablefootmark{b} & $\Delta v$\tablefootmark{c} & $N_\mathrm{ch}$\tablefootmark{d} & $S_\mathrm{peak}$\tablefootmark{e} & $S_\mathrm{tot}$\tablefootmark{f}\\
		& [mJy\,beam$^{-1}$\,ch$^{-1}$] & & [km\,s$^{-1}$] & & [Jy\,beam$^{-1}$] & [Jy] \\
		\hline
		\hline
		\multicolumn{6}{c}{VLBI} \\
		\hline
		Continuum    & 0.23 & $20\mathrm{mas}\times 20\mathrm{mas}$ & -    & 1   & $0.145\pm 0.020$ & $1.35\pm 0.20$\\
		Cube 	     & 0.37 & $20\mathrm{mas}\times 20\mathrm{mas}$ & 21.7 (43.4) & 143 & -	&	-   \\
		\hline
		\multicolumn{6}{c}{VLA}  \\
		\hline
		Continuum    & 2.1  & $1\farcs7\times 1\farcs2, -67\degr$ & -    & 1   & $2.883\pm 0.086$ & $3.37\pm 0.10$\\
		Cube (VLBI)   & 1.1  & $2\farcs0\times 1\farcs5, -81\degr$ & 21.7 (43.4) & 157 & -    & -   \\
		Cube (WSRT)  & 1.1  & $2\farcs0\times 1\farcs5, -81\degr$ & 20.0 (40.0) & 186 & -    & -   \\
		\hline
	\end{tabular}
	\tablefoot{ 
		\tablefoottext{a}{Noise level. For the cubes, this value represents the average value over all channels}
		\tablefoottext{b}{Synthesized beam as major axis, minor axis and the position angle}
		\tablefoottext{c}{Channel width in velocity of the cubes after re-sampling. The value in brackets corresponds to the effective resolution after Hanning-smoothing}
		\tablefoottext{d}{Number of channels of the cubes}
		\tablefoottext{e}{Total flux density of the continuum image}
		\tablefoottext{f}{Peak flux density of the continuum image}
	}
\end{table*}
\section{Observation \& Data reduction}
\label{sec:Obs}

	\subsection{VLBI Observation}
	\label{sec:Obs:VLBI}
	
		\object{\mbox{3C\,236}} was observed with a global VLBI array of 14 telescopes on 2015 Jun 13 (project code: GN002B). The observational setup is summarized in Table \ref{tab:Data:Observation}. The array included the full VLBA and stations from the European VLBI Network (EVN). In addition, Arecibo (Puerto Rico) participated for two hours. The observation lasted a total of 15 hours with the EVN and VLBA observing for $\sim 11$ hours and $\sim 12$ hours, respectively, and including an overlap between both arrays of $ \sim 8$ hours. The stations of Onsala and Kitt Peak (VLBA) were flagged during the calibration due to unusually high system temperatures and bandpass problems, respectively. The data were correlated at the Joint Institute for VLBI ERIC (JIVE) providing two data sets, i.e., one with 4 IFs each with 32 channels (`continuum pass') and one with one IF with 512 channels (`spectral-line pass'). \object{0958$+$346} (\object{BZQ\,J1001$+$3424}) was observed as the phase reference calibrator, while \object{0955$+$326} (\object{3C\,232}) and \object{0923$+$392} (\object{[HB89]\,0923$+$392}) served as the bandpass calibrators.
		
		The data were calibrated in two steps using standard procedures in the Astronomical Image Processing Software (\textsc{AIPS}, version 31DEC15) package \citep{AIPS1999} and beginning with the data from the continuum pass. The amplitude calibration and initial flagging were provided by the EVN pipeline. As a next step manual phase calibration on a single scan of the calibrator was performed to remove the instrumental delay. This was followed by a global fringe fit of the calibrators to correct for the phase delay and rate with the solutions applied to the target source. Finally, the bandpass was corrected using the bandpass calibrators and the solutions applied to the phase reference and target source.	For the spectral-line pass, the amplitude calibration and initial flagging was also taken from the EVN pipeline. The phase calibration was performed using the solutions from the manual phase calibration and global fringe fit of the continuum pass, which was followed by the bandpass calibration. Afterwards, the data were separated into a data cube with full spectral resolution and continuum data set with all channels averaged toghether. 
		
		The continuum data were further processed in \textsc{Difmap} \citep{Shepherd1994,Difmap2011}. This entailed imaging of the brightness distribution of the source using the \textsc{clean}-algorithm \citep{Hoegbom1974} in combination with phase self-calibration and flagging of corrupted visibilities. Once a sufficiently good model was found, a time-independent gain correction factor was determined for each telescope through amplitude self-calibration. The iterative process of imaging and phase self-calibration with subsequent time-dependent amplitude self-calibration was repeated several times with decreasing solution interval for amplitude self-calibration. 
		
		The resulting continuum image was used to perform a single phase self-calibration of the data cube with full spectral resolution in \textsc{AIPS}. After carefully inspecting the channels and flagging of corrupted visibilities, the continuum was subtracted in the visibility domain using a linear fit to the first and last 100 channels (\textsc{AIPS} task \textsc{UVLIN}). Since we focus on the faint and broad component of the absorption, we averaged over three consecutive channels to improve the sensitivity. The data were corrected for the Doppler shift in frequency caused by the rotation and movement of the Earth. Finally, a redshift correction was applied to the channel width in observed frequency to convert into rest-frame velocity following \cite{Meyer2017}. 
		
		Each channel of the spectral-line cube was imaged individually with robust weighting set to $1$ and a $(u,v)$-taper of 10\,M$\lambda$ to further improve the sensitivity. We found that this tapering of the visibility data provides the best combination of resolution and sensitivity. It is similar to \cite{Struve2012}. 
		The channels were only imaged if significant negative flux density was found in the area covered by continuum emission. The resulting cube covers a velocity range of $28117$--$31209\mathrm{\,km\,s^{-1}}$ at channel resolution of $21.7\mathrm{\,km\,s^{-1}}$ which is effectively doubled to $43.4\mathrm{\,km\,s^{-1}}$ due to Hanning-smoothing. 

		In order to compare the image cube and the continuum image, both were restored with the same circular restoring beam of 20\,mas. This is close to the synthesized beam due to the $(u,v)$-tapering and similar to \cite{Struve2012}. The noise levels were determined by fitting a Gaussian distribution to the pixels which are not contaminated by emission from the target following the procedure outlined in \cite{Boeck2012}. For the subsequent analysis the average noise level of the cube is used as a reference value. An overview of the image parameters is given in Table \ref{tab:Data:Image}. 
		The overall amplitude calibration uncertainty of the VLBI data was estimated to be around 15\% based on multiple iterations of imaging and self-calibration of the continuum data. The total flux density was measured in the image plane using the CASA Viewer \citep{CASA2011}. We estimated the uncertainty of the peak and total flux density measurement as $\sqrt{(N_\mathrm{beam}\times \sigma_\mathrm{noise})^2 + (0.15 \times S_\mathrm{tot})^2}$ following \cite{Nyland2016} where $N_\mathrm{beam}$ corresponds to the number of beams covered by the source. However, we noticed that in our case the first term has only a marginally impact on the uncertainty.

	\subsection{VLA Observation}
		The VLA observed \object{\mbox{3C\,236}} for 40\,min in A-array configuration (project code: 11A-166) on 2011 Aug 10. The setup is summarized in Table \ref{tab:Data:Observation}. \object{1328$+$307} (\object{3C\,286}) was used as the flux density calibrator and \object{1015$+$359} (\object{B2\,1015$+$35B}) as the complex gain calibrator.

		The data were processed with \textsc{Miriad} \citep{Sault1995,Miriad2011}. Due to technical difficulties one polarisation was lost and had to be flagged completely. The data were calibrated using standard procedures for the VLA. A continuum image was produced using \textsc{CLEAN} and self-calibration. The spectral-line data were continuum subtracted using only channels devoid of the broad absorption. Further processing was performed in \textsc{AIPS}. The VLA cube was averaged to match the velocity resolution of the VLBI data, cleaned in the region of the continuum emission and afterwards Hanning smoothed. Consistent with the VLBI spectral-line data, the channel width was redshift-corrected to the rest-frame velocity width. The resulting image cube covers a velocity range from $(28316$--$31706)\mathrm{\,km\,s^{-1}}$. For a comparison with the WSRT spectrum presented in \cite{Morganti2005}, we also created a second cube which matches the velocity resolution of the WSRT data. The parameters of the continuum image and the cube are given in Table \ref{tab:Data:Image}. We estimate the uncertainty of the absolute flux density scale at this frequency and given the calibrators to be $\sim 3\%$ based on \citep{Perley2013a}.

\section{Results}
\label{sec:Results}

	\subsection{VLA and VLBI continuum}
	\label{sec:Results:Cont}
		
		The VLA continuum image covers the central $6\arcmin\times6\arcmin$ ($650\mathrm{\,kpc}\times650\mathrm{\,kpc}$) of 3C\,236 (see Fig. \ref{fig:Collage}). The recovered radio emission is unresolved and yields a flux density of about $3.37\pm0.10\mathrm{\,Jy}$. No further extended emission is detected. The flux density is consistent with the value of $3.324\pm0.097\mathrm{\,Jy}$ obtained from the central region of the radio emission by the lower-resolution NVSS survey at 1.4\,GHz \citep{Condon1998}.
		The full VLBI continuum emission of \object{\mbox{3C\,236}} from our observation is shown in the bottom panel of Fig. \ref{fig:Collage}. The total flux density is about $1.35\pm\,0.20\mathrm{\,Jy}$ which is consistent with the value obtained by \cite{Struve2012} (1.36\,Jy). It corresponds to about 40\% of the flux density measured by the VLA observation. The source is significantly extended covering approximately 1\arcsec ($\sim 1.8\mathrm{\,kpc}$), but most of the radio emission is localized within 400\,mas ($\sim 720\mathrm{\,pc}$). 
		
		The difference in flux density between VLBI and VLA of about 2\,Jy can have different reasons. Firstly, the shortest baseline of the VLBI array limits the largest angular scale on which emission can be recovered to about 600\,mas. Any extended emission on larger scales is resolved out by the inteferometer. However, there could still be a significant amount of emission within the largest angular scale limit. Assuming the emission is uniformly distributed over the entire area, then an integrated flux density of more than 600\,mJy is necessary to achieve a brightness at the $3\sigma_\mathrm{VLBI,cont}$ sensitivity limit of the VLBI image. Secondly, the consistency between our VLA and the NVSS flux density shows that all of the undetected emission must be within the area covered by the synthesized beam of the VLA observation, i.e. $1\farcs7\times 1\farcs2$ ($3.1\mathrm{\,kpc}\times2.2\mathrm{\,kpc}$). Assuming again a uniform distribution of the undetected emission over this region leads to a brightness of about $0.6\mathrm{\,mJy\,beam^{-1}}$. This is just below the $3\sigma_\mathrm{VLBI,cont}$-limit. Thus, a global VLBI experiment including the VLA and eMERLIN would provide the sensitivity and short spacing to recover all of the emission.

		The morphology is overall consistent with previous VLBI observations by \cite{Schilizzi2001} and \cite{Struve2012}. The sensitivity of our VLBI image is about a factor of three better than the image from \cite{Struve2012}, but we do not detect significantly more extended emission or any movement of features in the jet at the given resolution. 
		Following \cite{Schilizzi2001}, we consider the location of the VLBI core region to coincide with the brightest feature at the phase center of the image and the jet to extend to the north-west. The emission in the south-east direction from the core region corresponds to the radio lobe created by the counter jet. Due to the chosen restoring beam, the brightest feature is actually a blend of the emission from the VLBI core and part of the innermost jet. We refer to this as the nuclear region from here on.
		In the following, we continue to focus on the inner most 400\,mas of the source where the bulk of the radio emission is located. We determine the position angle of the jet as the angle along which the brightest features of the south-east jet are best aligned on to be about 116\degr which is consistent with the measurement of \cite{Struve2012} of 117\degr. 


	\begin{figure}[t!]
		\centering
		\includegraphics[width=0.97\linewidth]{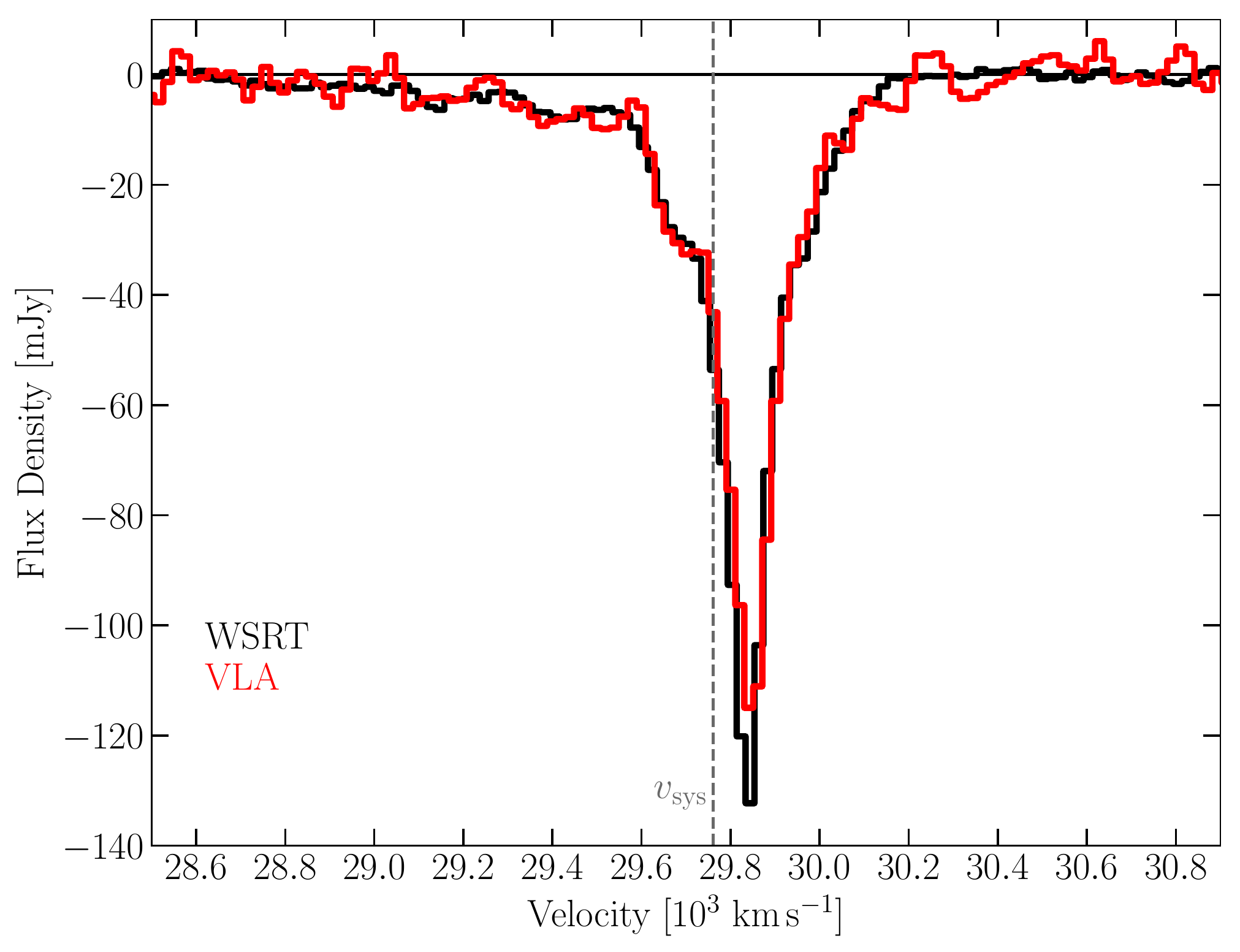}
		\includegraphics[width=0.97\linewidth]{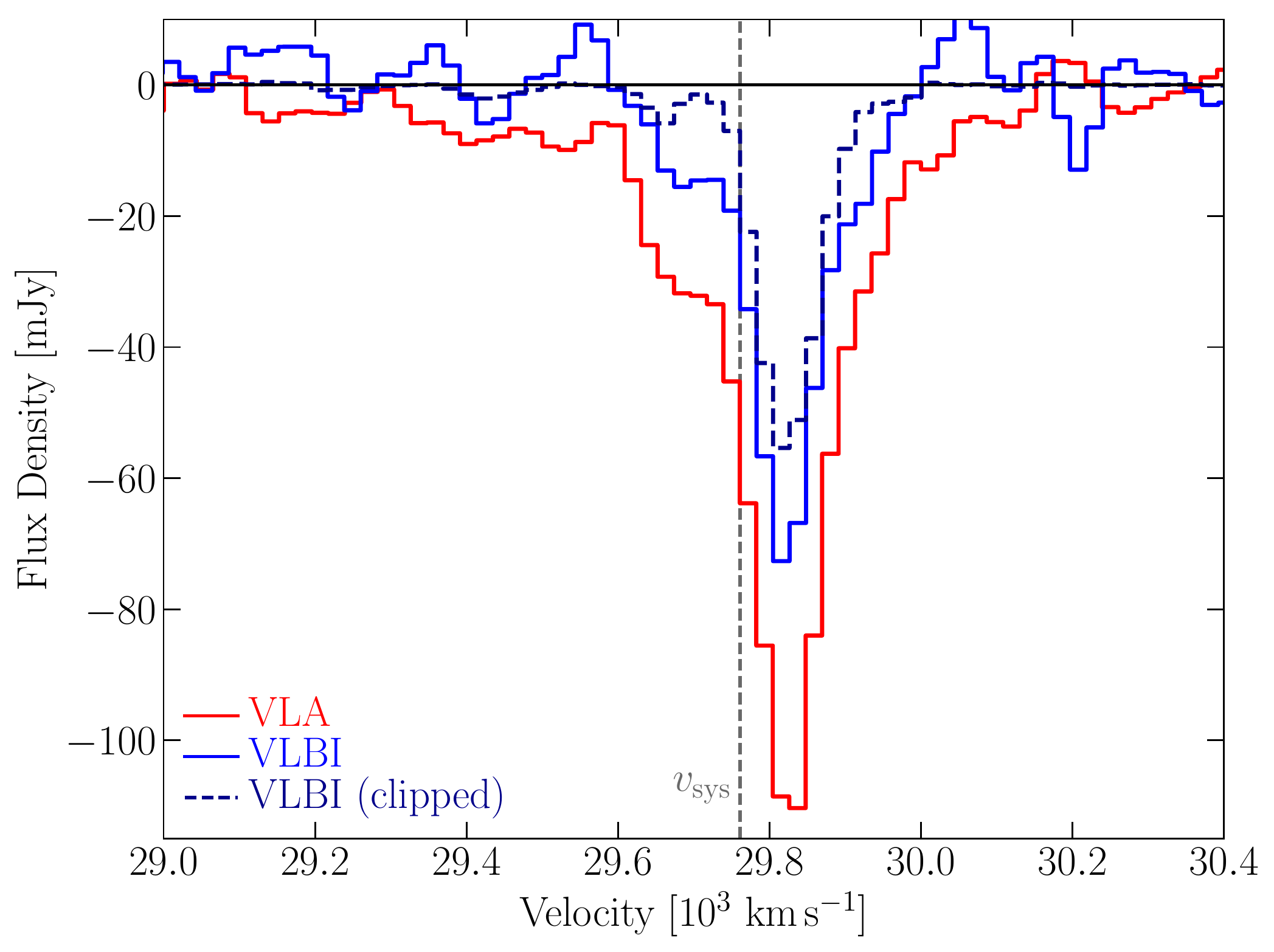}
		\includegraphics[width=0.97\linewidth]{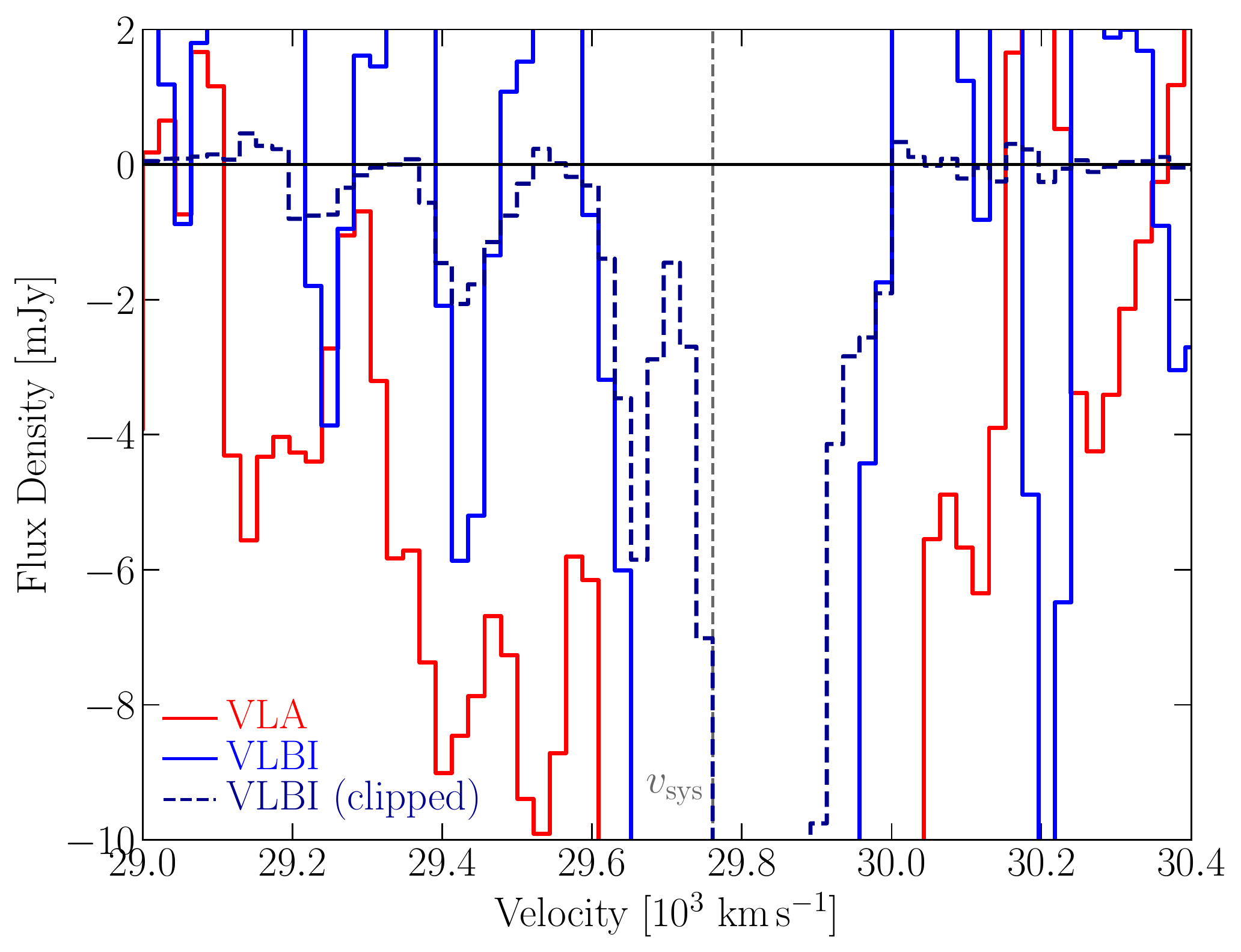}
		\caption{\ion{H}{I} absorption spectra of \object{\mbox{3C\,236}}. The dashed, vertical line marks the systemic velocity from \cite{Labiano2013}. Top panel: WSRT (black) spectrum from \cite{Morganti2005} and VLA (red) spectrum between $28500$--$30900\mathrm{\,km\,s^{-1}}$. Here, the velocity resolution of the VLA spectrum was matched to the WSRT data. Middle panel: The spatially integrated VLBI with (darkblue, dashed) and without (blue, solid) clipping of the cube pixels at the $3\sigma_\mathrm{VLBI,Cube}$ between $29000$--$30400\mathrm{\,km\,s^{-1}}$. For the VLA spectrum (red), the spectral resolution was matched to the VLBI spectrum. Bottom panel: Same as the middle panel, but zoomed-in in flux density.}
		\label{fig:Spec2}%
	\end{figure}

	\subsection{\ion{H}{I} absorption spectrum}
	\label{sec:Results:Spectrum}

		Figure \ref{fig:Spec2} (top panel) shows the unresolved VLA spectrum between $28500\mathrm{\,km\,s^{-1}}$ and $30900\mathrm{\,km\,s^{-1}}$ in combination with the WSRT  spectrum from \cite{Morganti2005}. The spectra taken with both instruments are consistent and show the same features, i.e., a deep and narrow absorption that smoothly falls off towards higher velocities, but has a complex, broad wing towards lower velocities. The consistency between both spectra implies that all of the absorption stems from scales smaller than the beam size of the VLA.

		Two spatially integrated VLBI spectra are shown in the bottom panel of Fig. \ref{fig:Spec2} between $29000\mathrm{\,km\,s^{-1}}$ and $30400\mathrm{\,km\,s^{-1}}$. Both were compiled by considering those pixels in the image cube that are located within the region marked by the $3\sigma_\mathrm{VLBI,cont}$ contour line of the continuum image. They differ in terms of the selection of pixels in the cube. For the dashed, blue line labelled `VLBI (clipped)' a very conservative limit of $|S_\mathrm{pixel,cube}|\geq 3\times\sigma_\mathrm{VLBI,cube}$ was used where $|S_\mathrm{pixel,cube}|$ is the absolute value of the pixel brightness. No such limit was applied for the compilation of the spectrum marked by the solid, blue line  (labelled `VLBI') and the VLA spectrum (solid, red line) which is unresolved in contrast to the VLBI spectrum. 
		
		The VLBI without clipping shows a deep and narrow absorption feature consistent with the previous measurement by \cite{Struve2012}, but its depth and width does not match the VLA reference spectrum. Because the deep absorption is likely due to gas associated with the extended dust lane (see Sect. \ref{sec:Results:GasDistribution}), the undetected absorption flux density is likely related to structure resolved out as seen from the missing continuum flux density. However, some unsettled gas appears in our VLBI observations (see Sect. \ref{sec:Results:GasDistribution}). More interesting, the observations reveal some of the outflowing gas (bottom panel of Fig. \ref{fig:Spec2}). However, the VLBI observation recover only a small fraction of the blue-shifted wing of the \ion{H}{I} profiles. We will discuss the possible implications of this for the distribution of the outflowing gas in Sect. \ref{sec:Discussion:outflow}.
	
	\begin{figure*}
		\centering
		\includegraphics[width=.95\linewidth]{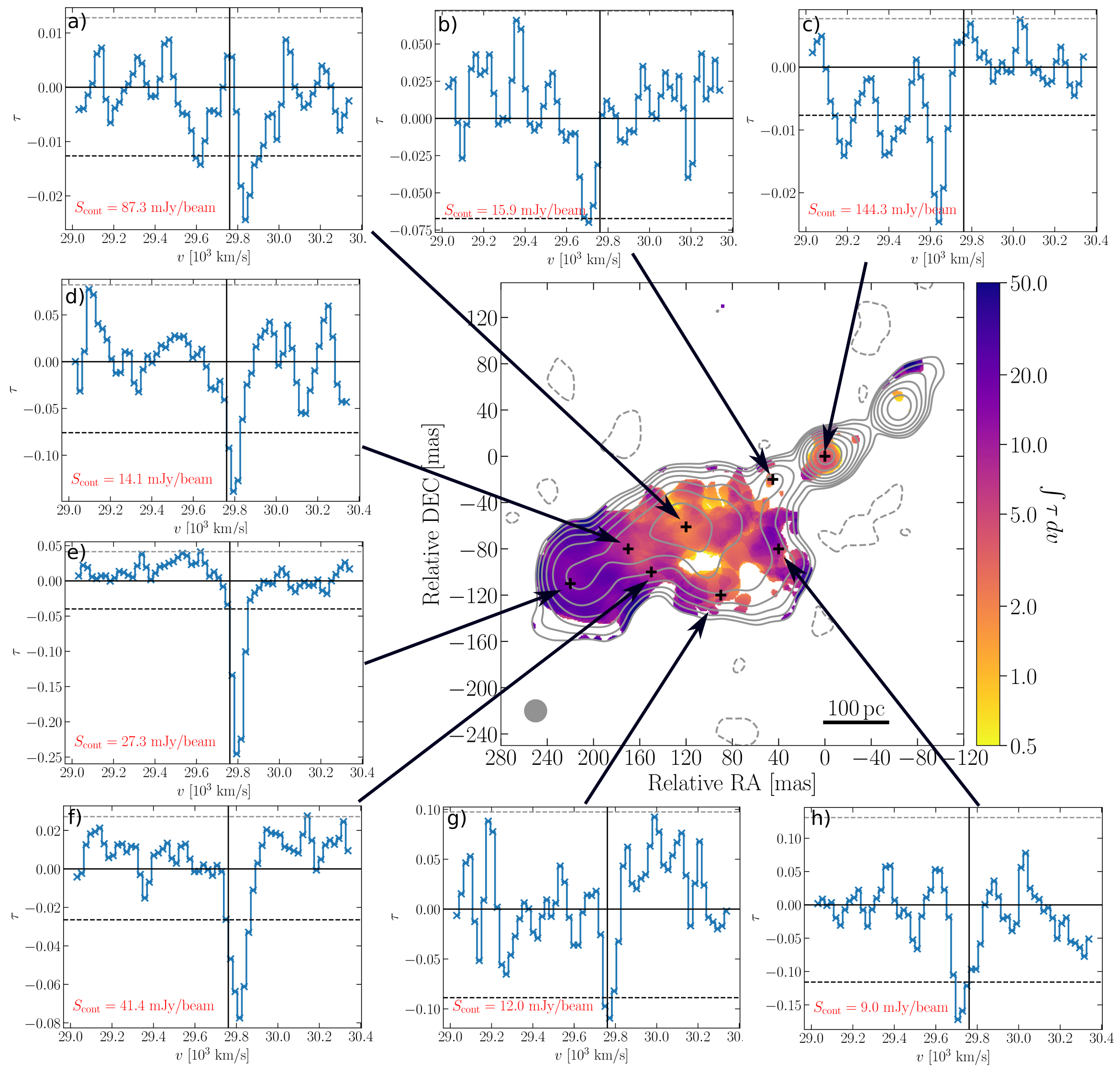}
		\caption{Central panel: Continuum image of the central region of \object{3C\,236} shown in Fig. \ref{fig:Collage}. The contour lines start at 3$\sigma_\mathrm{VLBI,cont}$ and increase logarithmically by a factor of 2. Solid and dashed contour lines correspond to positive and negative flux density, respectively. The black crosses mark areas where the spectrum was extracted between 29000--30000\,km\,s$^{-1}$. Outer panels: The optical depth spectrum shown as the ratio of the absorbed flux density to the continuum flux density. The dashed line represents the optical depth detection limit at the 3$\sigma_\mathrm{VLBI,cube}$-noise level. The solid, vertical line marks the systemic velocity.}
		\label{fig:tau}%
	\end{figure*}

	\subsection{\ion{H}{I} gas distribution}
	\label{sec:Results:GasDistribution}	

		The spatial distribution of the \ion{H}{I} gas is shown in the central panel of Fig. \ref{fig:tau}. It shows the optical depth $\tau$ integrated over the same velocity range as in the bottom panel of Fig. \ref{fig:Spec2}, i.e., between $29000\mathrm{\,km\,s^{-1}}$ and $30400\mathrm{\,km\,s^{-1}}$. The optical depth is defined as $\tau=\log(1-\Delta S_\mathrm{abs}/(c_f S_\mathrm{cont}))$ where $\Delta S_\mathrm{abs}$ and $S_\mathrm{cont}$ correspond to the absorbed and the continuum flux density, respectively, and the covering factor $c_f$ is assumed to be unity. In order to avoid integrating over noise and to get a reliable albeit conservative distribution of the \ion{H}{I} gas, we take into account only channels with $\leq-3\sigma_\mathrm{VLBI,cube}$. In addition to the integrated optial depth $\int\tau dv$, this figure shows single-pixel spectra of optical depth extracted at specific locations of the radio continuum, for which the detection limit of channels was not applied.

		The map of $\int\tau dv$ reveals a complex gas distribution across the south-east lobe and compact absorption towards the nucleus. In particular, the gas covering the south-east lobe exhibits significant changes in $\int\tau dv$. In the central and brightest part of the lobe $\int\tau dv$ reaches its lowest values, but there are several gaps in the gas distribution, in particular towards the nuclear region. The highest values of $\int\tau dv$ are measured towards the end of the south-east lobe. This is the region where the majority of the \ion{H}{I} gas leading to the narrow, deep feature in the integrated spectrum, is situated.

		The spectra of $\tau$ in Fig. \ref{fig:tau}a--h shows the range of optical depth probed by our observation. The lowest optical depth is reached towards the nuclear region as it has the brightest part of the radio source (Fig. \ref{fig:tau}c). Here, we are sensitive down to $\tau\gtrsim 0.0077$ at the $3\sigma_\mathrm{VLBI,cube}$-noise level and find three distinct kinematic features which will be discussed in greater detail later in this section. Towards the south-east lobe, the optical depth sensitivity varies. The lowest optical depth is reached in the central region of the lobe (Fig. \ref{fig:tau}a) with $\tau\gtrsim 0.013$ where we also find a more complex kinematic structure than in the other regions.

\begin{figure*}
		\centering
		\includegraphics[width=.84\linewidth]{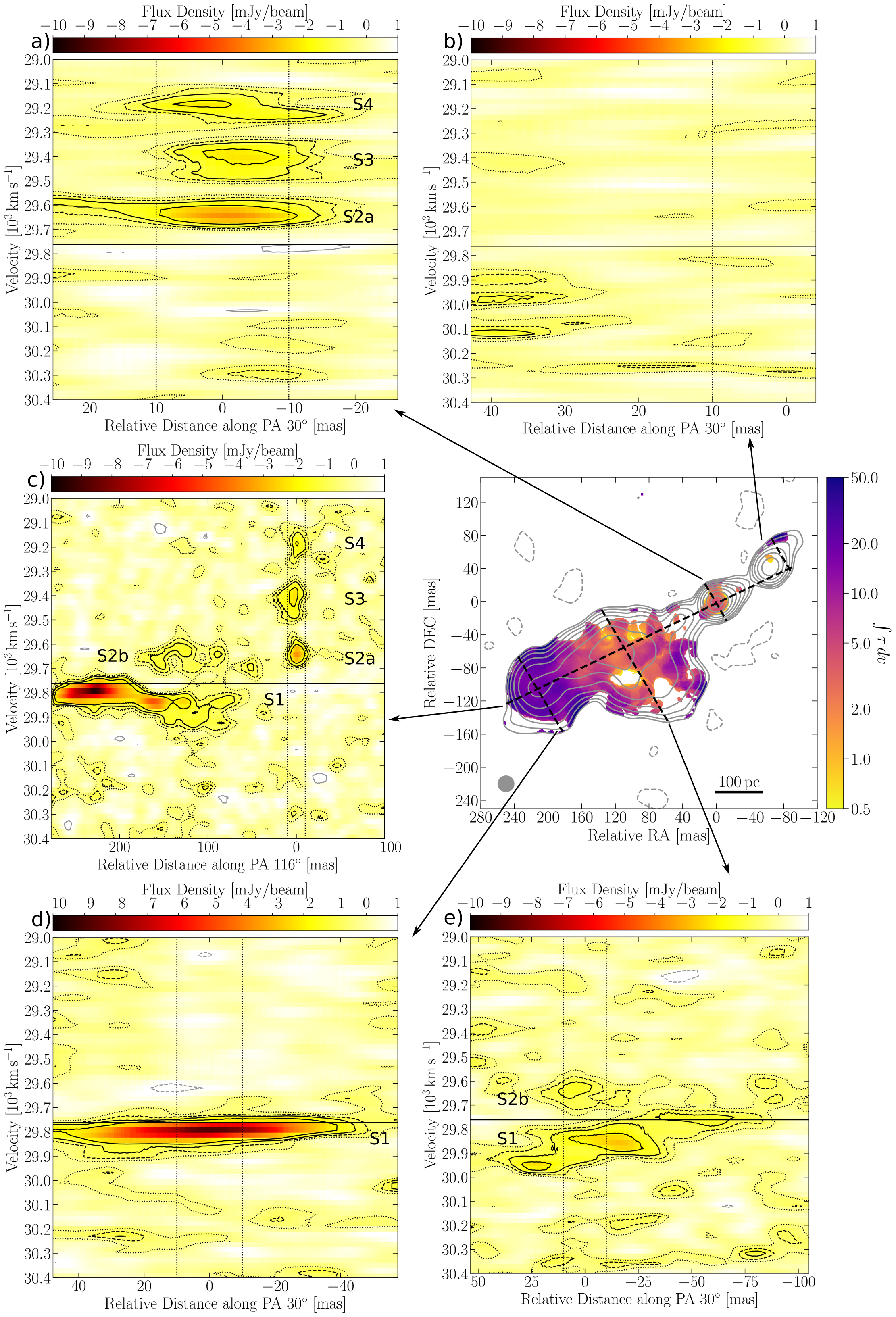}
		\caption{Position-velocity plots (panels a to e) from 29000--34000\,km\,s$^{-1}$ for slices along different position angles. The same colour scale is applied to all position-velocity plots. The black dashed vertical lines mark the size of the synthesized beam for comparison, while the black solid horizontal line refers to the systemic velocity from \cite{Labiano2013}. The black contour lines correspond to -1 (dotted lines), -2 (dashed lines), -3 and -5 (solid lines) times $\sigma_\mathrm{VLBI,cube}$, while the solid gray contour line is set to $3\sigma_\mathrm{VLBI,cube}$. The central panel shows the slices and the optical depth integrated over the same velocity range as the position-velocity plots, but including only channels with $\leq-3\sigma_\mathrm{VLBI,cube}$.}
		\label{fig:PV}%
	\end{figure*}

		In order to investigate the spatial and velocity distribution of these features in greater detail, we show position-velocity plots along different position angles in Fig. \ref{fig:PV}a--e. Again, we focus on the velocity range of $29000\mathrm{\,km\,s^{-1}}$ and $30400\mathrm{\,km\,s^{-1}}$. For comparison, the central panel shows the same map of $\int\tau dv$ as in Fig. \ref{fig:tau}.

		Figure \ref{fig:PV}c and d show the velocity structure of the gas along the jet position angle and along the position angle of the inner dust lane, respectively. In particular, Fig. \ref{fig:PV}d reveals a gradient in velocity similar to \cite{Struve2012} which has been interpreted as a signature of an \ion{H}{I} disk aligned with the inner dust lane. Our new data shows that this gradient is even more prominent across the central part of the lobe (Fig. \ref{fig:PV}e). At the $-2\sigma_\mathrm{VLBI,cube}$-level, the gradient would cover almost 300\,km\,s$^{-1}$. We label the disk-related feature in the following as S1. 
		
		However, more interesting and relevant from these observations are structures that appear to have a disturbed kinematics (see Figure \ref{fig:PV}c). S2b is located towards the central part of the lobe and is slightly blue-shifted with respect to the peak of the absorption by about $150\mathrm{\,km\,s^{-1}}$. At the $3\sigma_\mathrm{VLBI,cube}$-level it is not connected to S1 spatially or in velocity. The clear separation in velocity between S2b and S1 could indicate that S2b traces a different component of the \ion{H}{I} gas than S1.

		The most important result from the new observations is the finding that the gas co-spatial to the nucleus is entirely blue-shifted with respect to the peak absorption by up to $\sim 640\mathrm{\,km\,s^{-1}}$ (Fig. \ref{fig:PV}a,c). The projected size of this region is only about 36\,pc or even smaller. It comprises three distinct features labelled S2a, S3, and S4, which are separated in velocity by a few channels. While it is possible that this separation could be due to changes in sensitivity across the channels, we consider this to be the least likely explanation as the variation in sensitivity are not significant. This leaves three other possibilities (see also Sect. \ref{sec:Discussion:outflow}). First, there is no further \ion{H}{I} gas in this region or the optical depth of the gas is too low to be detectable. In this case the majority of the remaining ouflowing gas would have to be located elsewhere. Second, the cold \ion{H}{I} gas clouds could be entrained by warmer \ion{H}{I} gas which has a higher spin temperature and thus, a lower optical depth. Third, the gas within the outflow is highly clumpy which could imply differnces in the covering factor or spin temperature of the gas.

		We cannot exclude the possibility of \ion{H}{I} gas co-spatial to the north-west part of the jet. Figure \ref{fig:PV}b and the map of $\int\tau dv$ suggest that there could be gas in this region that is redshifted with respect to the deep absorption. However, these features are just at the $3\sigma_\mathrm{VLBI,cube}$-level, very narrow and generally located at the edge of the continuum emission. Therefore, we cannot consider them reliable detections.
				
\section{Discussion}
\label{sec:Discussion}

	Our VLBI observation has successfully recovered part of the outflowing \ion{H}{I} gas in 3C\,236 in the form of distinct, compact clouds (S2a, S3, S4). They are located primarily co-spatial, in projection, to the nuclear region which has a projected size of $\lesssim 36\mathrm{\,pc}$ with one possible exception (S2b). The clouds cover velocities of $150\text{--}600\mathrm{\,km\,s^{-1}}$ blue-shifted with respect to the \ion{H}{I} that is likely related to a rotating disk aligned with inner dust lane (S1). The disk-related gas extends over most over the south-east radio lobe. The overall gas distribution is clumpy with the majority of the gas concentrated at the end of the lobe. An important characteristic of the \ion{H}{I} VLBI data is the significant amount of \ion{H}{I} absorption that is not detected, but inferred from low-angular resolution observations with the VLA and the WSRT. 	
			
	\subsection{The high-velocity \ion{H}I gas}
	\label{sec:Discussion:outflow} 
	
		As mentioned in Sect. \ref{sec:Intro}, \object{\mbox{3C\,236}} is classified as a LERG, which makes it less likely that the AGN is able to produce powerful radiative winds which could couple to the dust of the galaxy and create strong gaseous outflows. This in turn leaves the radio jet as the main driver for the \ion{H}{I} outflow, i.e., features S2a, S3, S4, and perhaps S2b. However, we cannot exclude the possibility of S2b being part of the \ion{H}{I} disk (see also Sect. \ref{sec:Discussion:disk}).

		The location of S2a, S3, and S4 suggests that the outflow starts already very close to the nucleus of \object{\mbox{3C\,236}} (see also Fig. \ref{fig:nh}a, b, and c). As these structures are unresolved, we use the restoring beam as an upper limit of the extent of the \ion{H}{I} gas ($\lesssim36\mathrm{\,pc}$ in projection). In contrast, S2b is extended and we estimate its projected size to be about $54\mathrm{\,pc}\times 15\mathrm{\,pc}$.

		There are several implications for the undetected absorbed flux density in our VLBI observation. The diffuse extended continuum emission that is resolved out by the high-resolution of VLBI (see Section \ref{sec:Results:Cont}) limits the area over which we can probe for absorption. However, we could still be able to detect \ion{H}{I} absorption in regions outside of the detected continuum emission if the absorption is compact. In fact, in some channels the absorption does extend marginally outside of the $3\sigma_\mathrm{VLBI,cont}$ contour lines, but is below the $3\sigma_\mathrm{VLBI,cube}$-level.

		Another possibility is related to the small fraction of the outflow detected only against the nuclear region. This region is the brightest one with a peak flux density of about $145\mathrm{\,mJy}$ which corresponds to an optical depth limit of 0.0026 at the $1\sigma_\mathrm{VLBI,cube}$-level or 0.0077 at the $3\sigma_\mathrm{VLBI,cube}$-level. It is worth noting that some of the channels between S4 and S3 and between S3 and S2a have optical depths above the $1\sigma_\mathrm{VLBI,cube}$-level at the location of the peak continuum flux density. However, the peak flux density of the lobe is about $87\mathrm{\,mJy}$, which corresponds to an optical depth limit of 0.0043 at the $1\sigma_\mathrm{VLBI,cube}$-level or 0.013 at the $3\sigma_\mathrm{VLBI,cube}$-level. The peak optical depth of the clouds S3 and S4 is about 0.014 which is consistent with the $3\sigma_\mathrm{VLBI,cube}$-level within in our amplitude calibration uncertainty. Thus, we cannot entirely exclude the existence of clouds such as S3 and S4 at larger distances from the nucleus, i.e., agains the lobe. However, we cannot detect these clouds because of sensitivity limits. 
		
		This could be a likely situation and entails that the outflow is actually more extended but made up of clouds similar to the one detected against the nucleus. We cannot exclude a component of diffuse gas, but this would be also too faint to be detectable. We can exclude the presence of single clouds producing a depth of the absorption similar to the one detected by the VLA and WSRT (see Fig. \ref{fig:Spec2}). Taking the peak flux density of the lobe and the lowest absorption recovered by the WSRT at around $28800\mathrm{\,km\,s^{-1}}$ would correspond to an optical depth of 0.02. Therefore, such clouds would have been detected at least against the brighter part of the lobe. We would also not have seen this cloud towards the nuclear region, the peak flux density implies an optical depth of about 0.015 similar to S3 and S4.
		
		\subsubsection{Properties of the outflowing \ion{H}{I} gas}

\begin{figure*}
				\centering
				\includegraphics[width=1\linewidth]{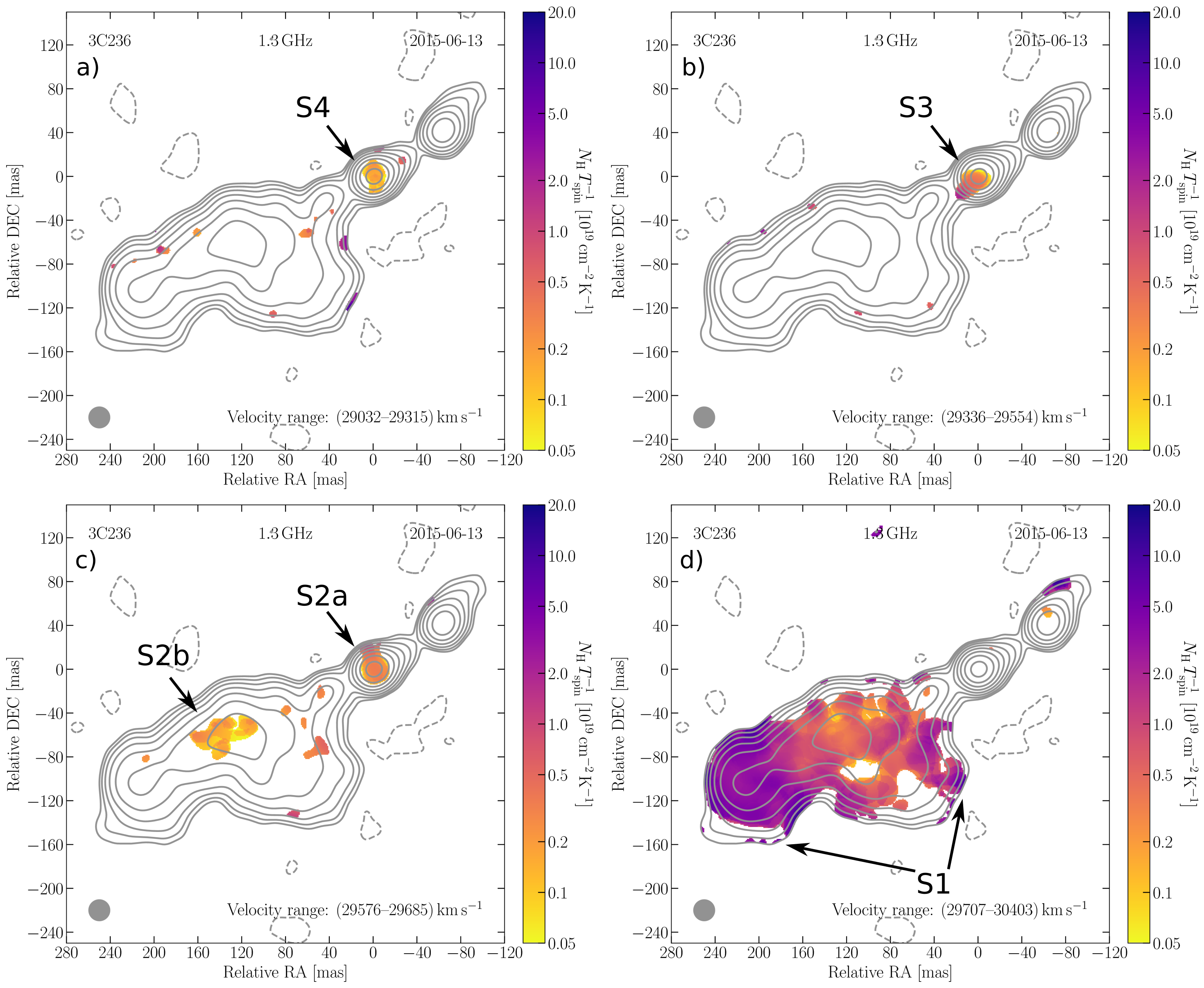}
				\caption{Maps of $N_\mathrm{\ion{H}{I}}T_\mathrm{spin}^{-1}$ for the features S1, S2a/b, S3 and S4 with the same logarithmic colour scale applied to all images. The velocity range is given in the bottom right corner and was chosen based on velocity range of the features in Fig. \ref{fig:PV}. The background shows the continuum map with contours beginning at 3$\sigma_\mathrm{VLBI}$ and increasing logarithmically by a factor of two. The grey circle in the lower left corner depicts the synthesized beam}
				\label{fig:nh}%
			\end{figure*}

			Figure \ref{fig:nh} shows the column density $N_\mathrm{H}$ normalized by the spin temperature $T_\mathrm{spin}$ following $N_\mathrm{\ion{H}{I}}T_\mathrm{spin}^{-1} \approx 1.823\times 10^{18} \int\tau(v)dv \mathrm{\,cm^{-2}\,K^{-1}}$, where $v$ is in units of $\mathrm{km\,s^{-1}}$. Again, we focus on the velocity range of $\sim29000\mathrm{\,km\,s^{-1}}$ to $\sim30400\mathrm{\,km\,s^{-1}}$ considering only channels with $\leq -3\sigma_\mathrm{VLBI,cube}$.
			The spin temperature of the \ion{H}{I} gas is unknown in \object{3C\,236}. However, it is very likely that it reaches higher values in the nuclear region than in the south-east lobe. A similar argument was made by \cite{Morganti2005b} and a $T_\mathrm{spin}$ of $\sim 1000\mathrm{\,K}$ is often used for gas close to the active nucleus or gas with extremely disturbed kinematics, although measurements of this quantity are scarce (see \citealt{Morganti2016,Holt2006}). Thus, it seems reasonable to assume $T_\mathrm{spin}=1000\mathrm{\,K}$ for S2a, S3 and S4 and $T_\mathrm{spin}=100\mathrm{\,K}$ for S2b. This results in column densities of $1.8\text{--}4.0\times 10^{21}\mathrm{cm^{-2}}$ for the clouds co-spatial to the nucleus and $1.8\times 10^{20}\mathrm{cm^{-2}}$ for S2b (see Table \ref{tab:Data:Outflow}). 

			We can further estimate the mass outflow rate $\dot{M}_\mathrm{\ion{H}{I}}$ following \cite{Heckman2002}

			\begin{align}
				\dot{M}_\mathrm{\ion{H}{I}} \sim 30 \frac{r_\star}{\mathrm{kpc}}\frac{N_\mathrm{\ion{H}{I}}}{10^{21}\mathrm{\,cm^{-2}}}\frac{v}{300\mathrm{\,km\,s^{-1}}}\frac{\Omega}{4\pi}M_\sun\mathrm{\,yr^{-1}}
			\end{align}
			
			where $r_\star$ is the distance of the cloud, $v$ its velocity and $\Omega$ its solid angle which is assumed to be $\pi$. We estimate $v$ for each of the features relative to the peak velocity of S1 (see Fig. \ref{fig:PV}c) and assume that $r_\star$ is given by the size of the beam. The only exception is S2b for which we measure a distance of $\sim150\mathrm{\,mas}$ in Fig. \ref{fig:PV}c leading to $r_\star \sim 310\mathrm{\,kpc}$ (de-projected).
			The resulting values are listed in Table \ref{tab:Data:Outflow}. As both $N_\mathrm{H}$ and $r_\star$ are lower limits, the actual mass outflow produced by these features is likely to be higher. Due to the order-of-magnitude lower column density compared to the clouds co-spatial to the nucleus, S2b has the lowest value. This is also the case for the density of all four clouds which ranges from $30\text{--}50\mathrm{\,cm^{-3}}$ for S2a, S3 and S4 compared to $2\mathrm{\,cm^{-3}}$ for S2b (Table \ref{tab:Data:Outflow}. For the former clouds a spherical geometry was assumed and for the latter an ellipsoidal geometry.

			\cite{Morganti2005} estimated the total mass outflow rate to be $\sim47M_\sun\mathrm{\,yr^{-1}}$ based on the full extent of the unresolved \ion{H}{I} absorption spectrum ($v\sim 1500\mathrm{\,km\,s^{-1}}$) and assuming an homogeneous distribution of the gas over a radius of 0.5\,kpc. Thus, we consider this value an upper limit on the total mass outflow rate. Integrating over S2a, S3 and S4 yields $N_\mathrm{H}\sim 7.8\times10^{21}\mathrm{\,cm^{-2}}$ and $\dot{M}_\mathrm{\ion{H}{I}}\sim 5M_\sun\mathrm{\,yr^{-1}}$. This suggests that at least 10\% of the estimated total mass outflow rate is located close (in projection) to the nucleus.
			
			Following \cite{Holt2006}, the kinetic energy of an outflow is given by

			\begin{align}
				E_\mathrm{kin} \approx 6.34\times 10^{35}\frac{\dot{M}}{2}\left(v_\mathrm{out}^2 + \frac{\mathrm{FWHM}^2}{1.85}\right)
			\end{align}
			
			where $\mathrm{FWHM}$ is the full width at half maximum of the line. \cite{Labiano2013} fitted three Gaussian distributions to the \ion{H}{I} WSRT spectrum. One of the distributions described the broad, blue wing of the spectrum with the following parameters: $v_\mathrm{hel}=29474\mathrm{\,km\,s^{-1}}$ and $\mathrm{FWHM}\sim 1100\mathrm{\,km\,s^{-1}}$. Using these values and $\dot{M}_\mathrm{\ion{H}{I},max}\sim 47M_\sun\mathrm{\,yr^{-1}}$ yields $E_\mathrm{kin}\sim 9.4\times 10^{42}\mathrm{erg\,s^{-1}}$. The estimated mass of the central supermassive black hole of \object{3C\,236} is $\log\,m_\mathrm{SMHB}\approx8.5\mathrm{M_\sun}$ \citep{Mezcua2011} and implies that the maximum kinetic energy of the \ion{H}{I} outflow is up to 0.02\% of the Eddington luminosity. For the VLBI detected outflow, we take the integrated mass outflow rate ($\gtrsim 5M_\sun\mathrm{\,yr^{-1}}$), the central velocity of S3 and the full width at zero intensity (FWZI) of $500\mathrm{km\,s^{-1}}$ (Fig. \ref{fig:PV}c). This suggests that the outflow close to the nucleus has at least 4\% of the maximum kinetic energy. 
			
			Recent numerical simulations of jet interaction with an inhomogeneous multi-phase ISM have reached resolutions which allow a comparison with VLBI measurements (by e.g., \citealt{Wagner2012,Mukherjee2016,Mukherjee2017,Cielo2017b}). This allows us to qualitatively consider the implications for our measurements.
			In general, the simulations demonstrated the strong impact that powerful jets have on the velocity, temperature, pressure and density distribution of the ISM gas. 
			When the jet hits the ISM, the gas at the shock front of the jet is being accelerated to the highest velocities. While the jet continues to push through the ISM, the already accelerated gas moves outwards along and transversal to the jet axis forming a kind of expanding cocoon around the jet. This disrupts the ISM in particular in the proximity of the jet and decreases the overall density of the gas. At some point the jet breaks through the ISM and the gas primarily expands transversal to the jet axis. However, the expansion of the jet can also be halted if the jet power is too low and/or the density of the medium is too high. This can prevent the jet from pushing all the way through the ISM. 
			The velocities and densities of the clouds that we measure are within the range of values expected from these simulations. Given the properties of the clumpy \ion{H}{I} gas and the morphology of the radio emission in 3C\,236, it seems likely that we see the jet-\ion{H}{I} interaction in an already advanced stage in its evolution. However, projection effects and the undetected \ion{H}{I} gas make it difficult to assess whether the VLBI jet has already entirely pushed through the \ion{H}{I} gas. 

			A more quantitive comparison is difficult as most of the available simulations consider only the warm and hot ISM gas ($\gtrsim10^4\mathrm{\,K}$). An exception is the recent study by \cite{Mukherjee2018} of \object{IC\,5063} which traces the cold molecular gas down to $10^2\mathrm{\,K}$. In contrast to \object{3C\,236} the jet axis in \object{IC\,5063} is aligned with the disk. In this particular work, the simulations were able to reconstruct kinematic features of the cold gas as seen in observations by \cite{Morganti2015}. Numerical simulations like the one performed by \cite{Mukherjee2018} are essential to understand the interaction of the radio jet and cold ISM gas.		

	\subsection{The disk-related \ion{H}{I} gas}
	\label{sec:Discussion:disk}

		\cite{Struve2012} related the rather symmetric gradient around the deep absorption to a disk of \ion{H}{I} gas (see Fig. \ref{fig:PV}d). They did not consider the gas detected within a distance of $<200\mathrm{\,mas}$ from the nucleus to be connected to it due to the lack of spatial and kinematic structure. However, our observations reveal a velocity gradient also across the central part of the lobe (Fig. \ref{fig:PV}d). It is larger than the one at the location of the deep absorption (Fig. \ref{fig:PV}a). Because the absorption extends to the edges of the continuum and not all of the absorption is recovered, it is difficult to measure the full width of the disk. Thus, we do not calculate the properties of the disk at this point as it would require more detailed modelling that is beyond the scope of this work.
		
		The column density of the \ion{H}{I} gas changes significantly across the lobe. Figure \ref{fig:nh}d depicts the gas related to S1. \cite{Struve2012} reported a value of $N_\mathrm{\ion{H}{I}}\approx 6.1\times 10^{21}\mathrm{cm^{-2}}$ at the location of the peak of the absorption assuming a conservative value of $T_\mathrm{spin}=100\mathrm{\,K}$. This is similar to our measurements, but we find that there is variation over an order of magnitude across the lobe. The \ion{H}{I} column densities are up to an order of magnitude lower than the column density from CO estimated by \cite{Labiano2013}.

		Assuming that the \ion{H}{I} disk is aligned with the inner dust lane, we can estimate the height of the \ion{H}{I} disk using the extent of S1 in Fig. \ref{fig:PV}c. This yields $\gtrsim 200\mathrm{\,mas}$ or $\gtrsim360\mathrm{\,pc}$ in projection and has to be considered a lower limit due to the undetected \ion{H}{I} gas. \cite{Schilizzi2001} estimated an apparent inclination of the radio jet to the line of sight of $\sim 60^\circ$ based on the ellipticity of the host galaxy and assuming that the jets are perpendicular to the dust lanes. This yields a de-projected height of the \ion{H}{I} disk of $\gtrsim 420\mathrm{\,pc}$. The major axis of the inner dust line is about 1.8\,kpc in projected size \cite{ODea2001} and the CO disk extends up to about 1.3\,kpc. Assuming that the \ion{H}{I} disk has the radial extent of the CO implies a thick rather than a thin disk. \cite{Nesvadba2011} also suggested an ellipsoidal configuration instead of thin disk for the H$_2$ gas on larger scales.

		The distribution of the gas across the south-east radio lobe seen in our VLBI image and in \cite{Struve2012} in addition to the co-spatiality with the inner dust lane provides further support for the interpretation that the morphology of the lobe is, to some extent, the result of interaction between the jet and the dust lane \citep{ODea2001}. Such an interaction would affect the morphology and kinematics of the \ion{H}{I} disk. The location and kinematic properties of S2b (see Fig. \ref{fig:PV}c,d and Fig. \ref{fig:nh}d) could be a signature of this interaction instead of being related to outflowing gas. In this context, it is interesting that \cite{Labiano2013} required two Gaussian functions to fit the deep part of the absorption spectrum, a deep, narrow component ($v_\mathrm{hel}=29828\mathrm{\,km\,s^{-1}}$, $\mathrm{FWHM}\sim 80\mathrm{\,km\,s^{-1}}$) and a shallower, broader one ($v_\mathrm{hel}=29846\mathrm{\,km\,s^{-1}}$, $\mathrm{FWHM}\sim 300\mathrm{\,km\,s^{-1}}$). Further investigations are necessary and would require detailed numerical simulations of the interaction between jet and the cold ISM. 
		
	\begin{table*}[htpb]
		\centering
		\caption[]{Properties of the kinematically disturbed HI clouds}
		\label{tab:Data:Outflow}
		\begin{tabular}{ccccccccc}
			\hline
			Component\tablefootmark{a} & $N_\mathrm{\ion{H}{I}}T_\mathrm{spin}^{-1}$\tablefootmark{b} & $N_\mathrm{\ion{H}{I}}$\tablefootmark{c} & $d$\tablefootmark{d} & $n_\mathrm{\ion{H}{I}}$\tablefootmark{e} & $m_\mathrm{\ion{H}{I}}$\tablefootmark{f} & $v$\tablefootmark{g} & $r_\star$\tablefootmark{h} & $\dot{M}_\mathrm{\ion{H}{I}}$\tablefootmark{i}\\
			& [$10^{19}$\,cm$^{-2}$\,K$^{-1}$] & [$10^{19}$\,cm$^{-2}$] & [pc] &  [cm$^{-3}$] &[$10^4$\,M$_\sun$] & [km\,s$^{-1}$] & [pc] & [M$_\sun$\,yr$^{-1}$] \\
			\hline
			\hline
			S4 		& 0.18 & 18 & $\lesssim36$ & 30  & 0.65 & 640 & $\lesssim40$ & 1.2\\
			S3 		& 0.40 & 40 & $\lesssim36$ & 60  & 1.5  & 420 & $\lesssim40$ & 1.7\\
			S2a 	& 0.33 & 33 & $\lesssim36$ & 50  & 1.2  & 150 & $\lesssim40$ & 0.5\\
			Nucleus & 0.78 & 78 & $\lesssim36$ & 120 & 2.8  & 640 & $\lesssim40$ & 5\\
			\hline
			S2b 	& 0.18 & 1.8 & $56\times 15$ & 2   & 0.28 & 150 & 310 & 0.2\\
			\hline
		\end{tabular}
		\tablefoot{ 
			\tablefoottext{a}{Label of the kinematic component (see Fig. \ref{fig:PV}). The values for `Nucleus' were obtained after integrating over S2a, S3, and S4}
			\tablefoottext{b}{\ion{H}{I} column density normalised by spin temperature}
			\tablefoottext{c}{\ion{H}{I} column density. For S4, S3, and S2a $T_\mathrm{spin}=1000\mathrm{\,K}$ is assumed, while for S2b $T_\mathrm{spin}=100\mathrm{\,K}$ is assumed.}
			\tablefoottext{d}{Projected size of the components. For S2a, S3, and S4 a spherical geometry is assumed with an upper limit of diameter based on the synthesized beam. For S2b an ellipsoidal geometry is assumed with the major and minor axis given.}
			\tablefoottext{e}{Density of the \ion{H}{I} clouds. The same $T_\mathrm{spin}$-values as for the column density are assumed here.}
			\tablefoottext{f}{Mass of the \ion{H}{I} clouds for the chosen $T_\mathrm{spin}$-values.}
			\tablefoottext{g}{Peak velocity of the \ion{H}{I} clouds relative to the peak velocity of S1.}
			\tablefoottext{h}{De-projected distance of the \ion{H}{I} clouds relative to the nucleus}
			\tablefoottext{i}{Mass outflow rate following \cite{Heckman2002} for the chosen $T_\mathrm{spin}$-values.}
		}
	\end{table*}

	\subsection{Comparison with 4C\,12.50}
	\label{sec:Discussion:Comparison}
	
		As mentioned in Sect. \ref{sec:Intro} the compact radio galaxy \object{\mbox{4C\,12.50}} located at a redshift of 0.1217 is currently the only other powerful radio galaxy for which the \ion{H}{I} outflow has also been studied with VLBI. There have been other radio galaxies in which a strong \ion{H}{I} outflow has been detected and partially resolved, i.e, 3C\,305 \citep{Morganti2005b}, 3C\,293 \citep{Mahony2013}. However, these observations were obtained at lower angular resolution and thus probed larger spatial scales. Therefore, we focus our comparison on \object{\mbox{4C\,12.50}}.
		
		\cite{Morganti2013} show that the \ion{H}{I} gas is distributed on either end of its projected 200\,pc-size radio structure, i.e., the deep absorption is located at the northern extent of the source, while the broad outflow is co-spatial with the hot spot in the southern part of the source. In contrast to 3C\,236, no absorption was reported in the nuclear region and the high- and low-resolution \ion{H}{I} spectrum match well which suggests that all of the absorption has been recovered by the VLBI observation. 
		
		\cite{Morganti2013} measured a column density of the blue-shifted clouds in \object{\mbox{4C\,12.50}} of $4.6\times 10^{21}\mathrm{cm^{-2}}$, using  $T_\mathrm{spin}=100\mathrm{\,K}$ due to the distance of the \ion{H}{I} gas to the nucleus. This is comparable to S2a, S3 and S4 even though these are located co-spatial to the nucleus, i.e., a higher value for $T_\mathrm{spin}$ was assumed. The mass outflow rate of the \ion{H}{I} in \object{\mbox{4C\,12.50}} was determined to range between $16\,M_\sun\mathrm{\,yr^{-1}}$ and $29\,M_\sun\mathrm{\,yr^{-1}}$. However, it is difficult to compare with 3C\,236 as we only measure lower limits. Although S2b would be better suitable for comparison in terms of its location, its column density and total \ion{H}{I} mass is an order of magnitude lower than what was determined for \object{\mbox{4C\,12.50}}.
		The mass outflow rate and outflow velocity suggests a kinetic energy of about 0.02--0.03\% of the Eddington luminosity for \object{\mbox{4C\,12.50}} depending also on the assumed black hole mass \citep{Dasyra2006,Dasyra2011b,Son2012}. This value range is similar to what we have estimated as a possible upper limit for the kinetic energy of the outflow in \object{\mbox{3C\,236}}. There is also potentially a large difference in the total \ion{H}{I} mass. \cite{Morganti2013} estimated a mass of $\sim1.4\times 10^5M_\sun$ for \object{\mbox{4C\,12.50}} which represents a lower limit as the \ion{H}{I} gas could be distributed beyond the radio continuum. \cite{Struve2012} derived a value of 5.9--9$\times 10^9M_\sun$ assuming the \ion{H}{I} at the southern end of the lobe in \object{\mbox{3C\,236}} is contained within a regular rotating disk.

		The differences observed between these two objects can be the result of a combination of differences in size and age of the two sources and differences in the conditions of the ISM. The gas in 3C\,236 could be more settled than in 4C\,12.50. This may suggest that at the distances of the south-east radio lobe of 3C236 of from the nuclear region of about 0.3 kpc, there are no more dense clouds and the jet has already broken through the denser gas. This would open the possibility that both sources represent different stages of evolution at least with respect to the jet-\ion{H}{I} interaction. \object{\mbox{3C\,236}} could be further advanced in its evolution than \object{\mbox{4C\,12.50}}. 
		The age of both radio sources have been estimated based on cooling time to be $\sim10^4\mathrm{\,years}$ (\object{\mbox{4C,12.50}}, \citealt{Morganti2013}) and $\sim10^5\mathrm{\,years}$ (\object{\mbox{3C\,236}},\citealt{ODea2001,Tremblay2010}). The jet in 3C\,236 could have had more time to interact with the \ion{H}{I} dispersing the gas to greater extent. Thus, the combination of these parameters need to be considered when the presence (or absence) of outflows and their properties are investigated.

\section{Summary \& Conclusion}
\label{sec:Summary}
	In this paper, we have presented results on the \ion{H}{I} gas distribution in the central 1\,kpc of the radio source \object{\mbox{3C\,236}} as detected in absorption by milli-arcsecond global VLBI and arc-second VLA observations. We find that all of the \ion{H}{I} gas recovered by VLBI is contained within the nuclear region and the south-east lobe of the radio source. The VLBI data recovers a significant amount of absorption from the disk-related and the outflowing \ion{H}{I} gas component compared to the lower resolution VLA data as well as about 40\% of the continuum flux density. The latter implies substantial extended low-surface brightness emission that is resolved out by the high resolution of VLBI.
	
	For the first time, we have been able to localise part of the broad blue-shifted component of the \ion{H}{I} gas in \object{3C\,236} in the form of distinct clouds located almost exclusively in the compact nuclear region which is $\lesssim 40\mathrm{\,pc}$ in size (in projection). The clouds cover a velocity range of about 600\,km\,s$^{-1}$ with respect to the peak of the disk-related \ion{H}{I} gas and we estimate that they have a density of $\gtrsim 60\mathrm{cm^{-3}}$. There is also one cloud co-spatial to the south-east lobe and well aligned with the position angle of the jet that appears to be kinematically disturbed gas. While it could be part of the \ion{H}{I} disk, it is also possible that it resembles outflowing \ion{H}{I} gas. In the latter case, its location and extended size implies a density as low as $\sim 2\mathrm{cm^{-3}}$. Overall, the mass outflow rate of the VLBI detected outflowing gas is about 10\% of the total mass outflow rate of $47\mathrm{M_\sun\,yr^{-1}}$ estimated from unresolved spectra. The clouds co-spatial to the nuclear region account for about 4\% of the total kinetic energy of the \ion{H}{I} outflow. Because \object{3C\,236} is classified as a LERG, we consider the radio jets as the most likely driver of the outflow.

	The discrepancy between the low- and high-resolution \ion{H}{I} absorption spectra in combination with the distribution of the detected gas implies that both the observed and undetected \ion{H}{I} outflow is clumpy. However, we cannot exclude the possibility of a highly diffuse gas component. A qualitative comparison with numerical simulations suggests that the interaction of the jet with the \ion{H}{I} gas has already been going on for long time in \object{\mbox{3C\,236}}. In this scenario the high-velocity gas that we do not detect has been dispersed significantly as a result of the jet interaction. Even the disk-related \ion{H}{I} gas could have been affected.
	
	We compare our results \object{\mbox{3C\,236}} to \object{\mbox{4C\,12.50}}, in which no absorption was detected co-spatial to the nuclear region and all of the \ion{H}{I} was recovered by VLBI including the jet-driven outflowing gas. The differences to \object{\mbox{3C\,236}} are very intriguing as it could be a sign that the gas is more settled in \object{3C\,236}. However, the available data does not allow to draw strong conclusions on whether both sources represent different stages in AGN evolution. Additional data on other sources are required.
	
	This work is part of our ongoing effort to spatially resolve the jet-driven \ion{H}{I} outflow in young and re-started powerful radio galaxies. It demonstrates that great care is required when physical quantities such as density and mass of the gas are derived from unresolved spectra. This shows the need for high-resolution follow-up observations of upcoming large \ion{H}{I} absorption surveys conducted by e.g., Apertif \citep{Oosterloo2010,Maccagni2017}, MeerKat \citep{Gupta2017}, and ASKAP \citep{Allison2015}.
   
\begin{acknowledgements}
	RS gratefully acknowledge support from the European Research Council under the European Union's Seventh Framework Programme (FP/2007-2013)/ERC Advanced Grant RADIOLIFE-320745.
	EKM acknowledges support from the Australian Research Council Centre of Excellence for All-sky Astrophysics (CAASTRO), through project number CE110001020.
	The European VLBI Network is a joint facility of independent European, African, Asian, and North American radio astronomy institutes. Scientific results from data presented in this publication are derived from the following EVN project code: GN002. 
	The National Radio Astronomy Observatory is a facility of the National Science Foundation operated under cooperative agreement by Associated Universities, Inc. The Long Baseline Observatory is a facility of the National Science Foundation operated under cooperative agreement by Associated Universities, Inc.
	The Arecibo Observatory is a facility of the National Science Foundation (NSF) operated by SRI International in alliance with the Universities Space Research Association (USRA) and UMET under a cooperative agreement. The Arecibo Observatory Planetary Radar Program is funded through the National Aeronautics and Space Administration (NASA) Near-Earth Objects Observations program.
	Based on observations made with the NASA/ESA Hubble Space Telescope, and obtained from the Hubble Legacy Archive, which is a collaboration between the Space Telescope Science Institute (STScI/NASA), the Space Telescope European Coordinating Facility (ST-ECF/ESA) and the Canadian Astronomy Data Centre (CADC/NRC/CSA).
	This research has made use of NASA's Astrophysics Data System Bibliographic Services.
	This research has made use of the NASA/IPAC Extragalactic Database (NED) which is operated by the Jet Propulsion Laboratory, California Institute of Technology, under contract with the National Aeronautics and Space Administration.
	This research made use of Astropy, a community-developed core Python package for Astronomy \citep{Astropy2013}.
	This research made use of APLpy, an open-source plotting package for Python \cite{Aplpy2012}.
\end{acknowledgements}

%
%
\bibliographystyle{aa} 
\bibliography{References} 


%

\end{document}